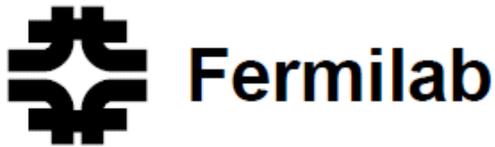

# Measurements and calculations of air activation in the NuMI neutrino production facility at Fermilab with the 120-GeV proton beam on target[*]

I. L. Rakhno[†], J. Hylen, P. Kasper, N. V. Mokhov, M. Quinn, S. I. Striganov, K. Vaziri

Fermi National Accelerator Laboratory, Batavia, Illinois 60510, USA

## Abstract

Measurements and calculations of the air activation at a high-energy proton accelerator are described. The quantity of radionuclides released outdoors depends on operation scenarios including details of the air exchange inside the facility. To improve the prediction of the air activation levels, the MARS15 Monte Carlo code radionuclide production model was modified to be used for these studies. Measurements were done to benchmark the new model and verify its use in optimization studies for the new DUNE experiment at the Long Baseline Neutrino Facility (LBNF) at Fermilab. The measured production rates for the most important radionuclides – $^{11}$C, $^{13}$N, $^{15}$O and $^{41}$Ar – are in a good agreement with those calculated with the improved MARS15 code.

[†]rakhno@fnal.gov

# Measurements and calculations of air activation in the NuMI neutrino production facility at Fermilab with the 120-GeV proton beam on target

I.L. Rakhno [*], J. Hylen, P. Kasper, N.V. Mokhov, M. Quinn, S.I. Striganov, K. Vaziri

* Corresponding author. Tel.: +1 630 840 6763, E-mail address: rakhno@fnal.gov

*Fermi National Accelerator Laboratory, Batavia, Illinois 60510, USA*

## Abstract

**Measurements and calculations of the air activation at a high-energy proton accelerator are described. The quantity of radionuclides released outdoors depends on operation scenarios including details of the air exchange inside the facility. To improve the prediction of the air activation levels, the MARS15 Monte Carlo code radionuclide production model was modified to be used for these studies. Measurements were done to benchmark the new model and verify its use in optimization studies for the new DUNE experiment at the Long Baseline Neutrino Facility (LBNF) at Fermilab. The measured production rates for the most important radionuclides – $^{11}$C, $^{13}$N, $^{15}$O and $^{41}$Ar – are in a good agreement with those calculated with the improved MARS15 code.**

## 1. Introduction

One of the most important radiation safety problems associated with the design and operation of high-power high-energy accelerators and their experiments is the release of radioactive gases into the atmosphere.  In particular, $^{41}$Ar with a relatively long half-life of 110 minutes is radiologically significant for off-site doses. One of the most challenging examples is neutrino experiments – planned (LBNF/DUNE) [1] and operational (NuMI/NOvA) [2] – with their 120-GeV proton, Megawatt scale beam power on target. To predict the air activation levels for a new facility, when direct measurements are not yet available, a combination of calculations and measurements from a similar existing facility are used. As an example, one can use measurements at the NuMI beam line, which has been operating at Fermilab for many years, to benchmark calculations used to predict air activation at future facilities.  This paper describes measurements of the radionuclides production rates in the NuMI target chase and compares them with MARS15 calculations.

Once a beam line or experimental facility starts operations, the air composition at the release points is analyzed and quantified to get an actual release estimate. However, before the operation of a beam line starts, an estimate of the levels radioactive gas emissions is needed as an input to the design of the facility.  Contemporary Monte Carlo codes such as MARS15 [3] provide quite reliable predictions for a wide range of physical quantities. Benchmarking such code predictions against experimental data provides a very useful information on the range of reliability regarding the air activation calculations. This study presents such an effort to measure the production of $^{41}$Ar and other important isotopes in the target chase at the NuMI beam line and compare the results with predictions by the MARS15 code.

When performing the air activation calculations, one faces the following problems: (i) primary and secondary particles of various types can contribute to the air activation; (ii) a very wide energy range should be taken into account (in our case spanning 14 decades, from 120 GeV down to thermal neutron energies); (iii) light target nuclei like oxygen, nitrogen and argon still represent a challenge for most of the cascade and evaporation modeling codes.

The general-purpose Monte Carlo code MARS15 allows modeling of particle and heavy ion interactions with matter from a multi-TeV region down to thermal neutron energies as well as spatial transport in arbitrary three-dimensional heterogeneous structures. To improve the quality of predictions for air activation levels, two upgrades were recently performed to MARS15 physics models: (i) production cross sections for a number of light target nuclei and most important projectiles were corrected in a very broad energy range; (ii) in the energy range from 1 up to 200 MeV, a completely new model of nuclear interactions, based on the TENDL library [4], was implemented for the most important projectiles.

The experimental and simulation procedures are described in the next two sections followed by comparisons between results of these measurements and MARS calculations.

## 2. Experimental setup and measurements

### 2.1. Experimental setup and air circulation model

Figure 1 shows a drawing of the NuMI target chase with target and horns, where the air samples were taken. The experimental setup is shown in Fig. 2. A corresponding fragment of the MARS model, that shows the carbon target and the magnetic focusing horn 1 filled with air, is shown in Fig. 3. Downstream of horn 1 there is another horn. The chase is surrounded by shielding, which altogether represents the NuMI target pile.

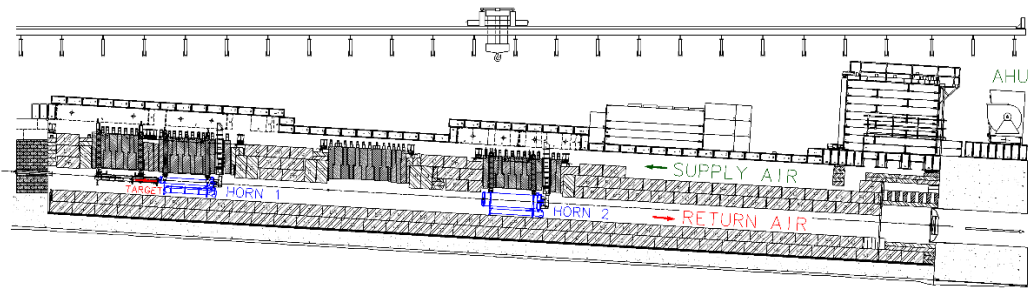

**Fig. 1.** A schematic diagram of the NuMI target pile showing the 120-cm long target and two focusing horns. This is the volume from which the air samples were taken. More detailed info on the target with its supporting structures, horns etc. is provided in Ref. [2] and in Fig. 3 below. The AHU refers to an air handling unit for recirculating air. The distance between upstream end of the target and AHU is approximately 50 m.

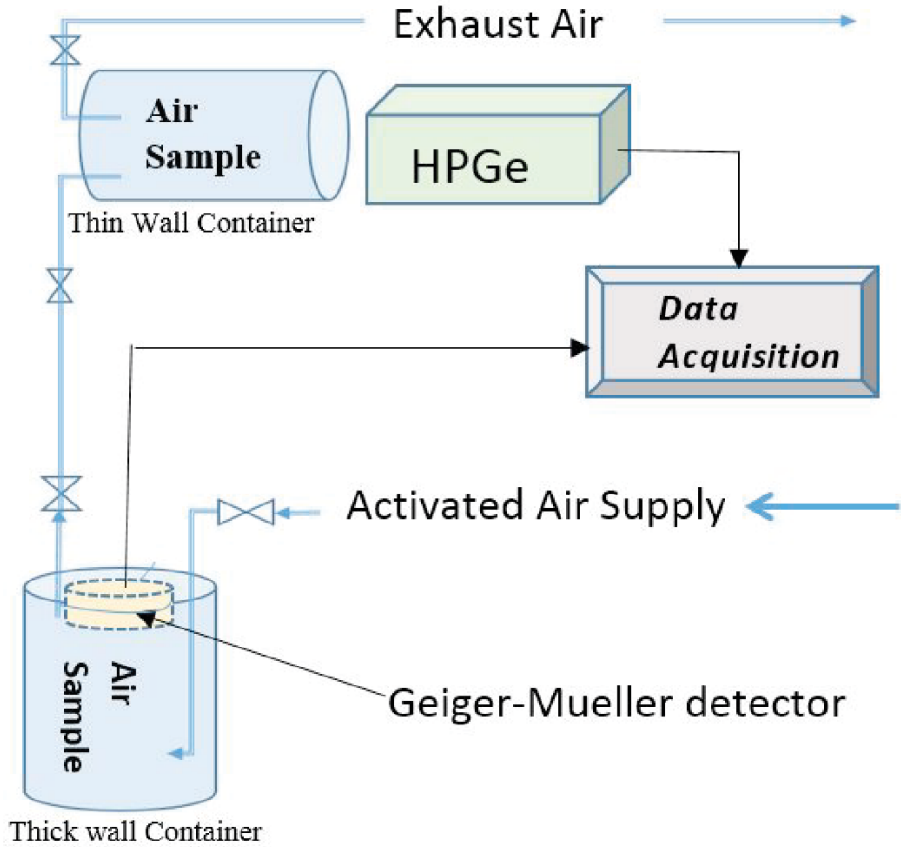

**Fig. 2.** A schematic drawing of the sampling and measuring configuration. Activated air from NuMI target chase was circulated through the sampling containers until the concentration levels reached an equilibrium. The valves shown in the supply and exhaust lines were then closed to isolate the samples before measuring their decay rates, "in situ," with the two different detector types: a High Purity Ge detector (HPGe), and a Geiger-Mueller counter (GM).

The activated air is circulated through the target chase by means of the target chase cooling system. A small amount of air continuously leaks out of the target chase air cooling loop. This leakage is taken into account by including an additional effective decay term, $\lambda_L$, which is the ratio of the air circulation rate and total air volume (see, e.g., [5, 6]). For this analysis, a complete air mixing model [6] is used. According to this model, an activated nucleus has the same probability of being removed from the region where the air activation and complete mixing occurs, no matter where it is produced. The master equation for this model, which describes temporal dependence of the number of radionuclides in a region of interest, is the following:

$$\frac{dN_i}{dt} = P_i v_p - (\lambda_i + \lambda_L) N_i, \qquad (1)$$

where $N_i$ is the number of radionuclides of the certain type i in this region, $P_i$ is the radionuclide production per proton on target (POT), $v_p$ is the properly normalized incident proton rate, and $\lambda_L$ is the correction to the decay constant, $\lambda_i$, due to the air leak described above.

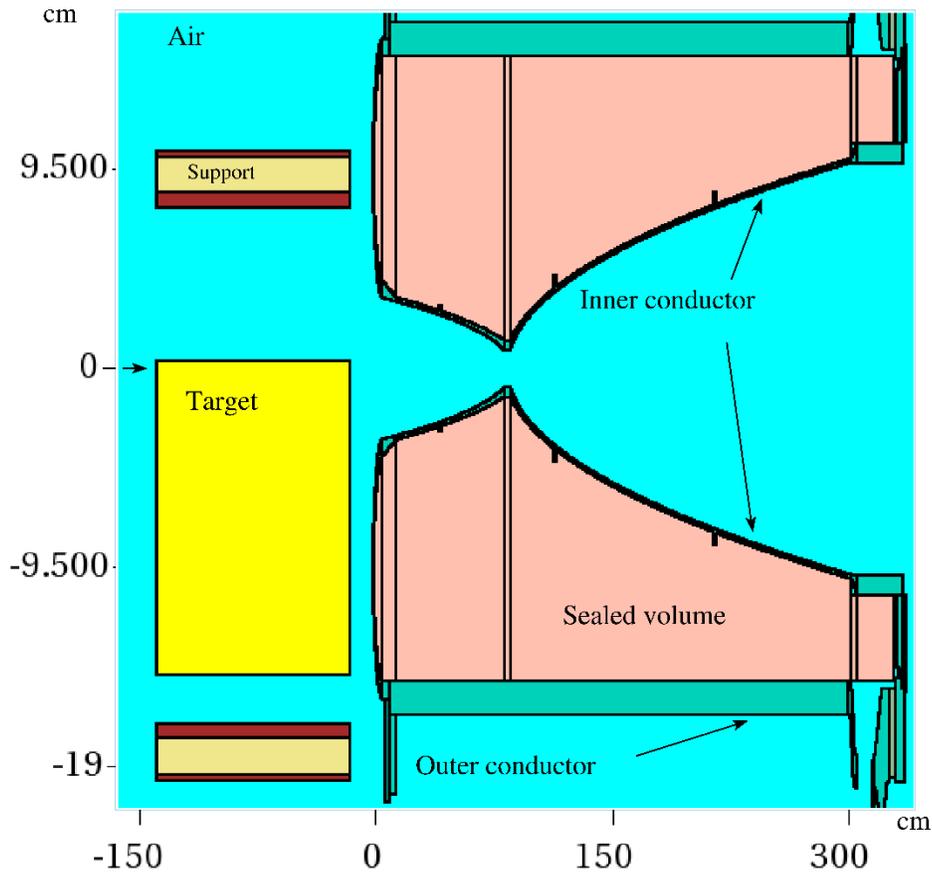

**Fig. 3.** A fragment of the MARS15 model (elevation view) that shows the pion-producing carbon target (rectangular volume shown in yellow), support and cooling structure around the target, and horn 1. In horn1, the sealed volume between the inner and outer conductors is filled with an inert gas. The arrow represents the incident 120-GeV proton beam. The dimensions are given in cm. Air filled areas are shown in cyan.

*2.2. Analytical description of radionuclides build-up and cool-down*

To compare activation measurements and predictions, we will need to account for actual irradiation and cooling temporal profile. Here we will introduce the formulae that are used for this purpose. The build-up of radioactive isotopes per unit volume produced during a period of constant irradiation is described by the well-known formula:

$$N_b(t)/V = \varphi\rho\sigma\lambda^{-1}\left(1-e^{-\lambda t}\right) = P\lambda^{-1}\left(1-e^{-\lambda t}\right) \qquad (2)$$

In Eq. (2) φ is the flux of particles produced by projectiles with energy above the corresponding threshold. In the standard methodology employed at Fermilab, this is defined as the flux of hadrons with kinetic energy above 30 MeV averaged over the target air volume, V. This is usually obtained from a MARS simulation in units of particles/cm$^2$/proton and is then scaled to the expected number of protons per second. The target density, $\rho$, is in units of atoms/cc and $\sigma$ is the production cross section in cm$^2$. The variable $\lambda$ is the isotope's decay constant in units of s$^{-1}$.

The first three terms in Equation (2) define the normalized isotope production rate P that measures the number of isotopes produced per primary beam proton on target (POT). To calculate the total number of isotopes produced one need simply scale the result of the equation by the volume and beam rate. Note that if the irradiation time is large with respect to the isotope's lifetime, it reaches a saturation concentration of P/$\lambda$ and a saturation activity per unit volume equal to the production rate, P.

The reduction during a cool-down period is simply given by the exponential decay formula:

$$N_c(t) = N_0 e^{-\lambda t} \qquad (3)$$

Equations (2) and (3) can be combined to determine the isotopes remaining after a sequence of irradiation or cool-down periods. This is done by simply adding the number of isotopes produced in each period from 1 to n weighted with cool-down terms for the earlier ones (with $T_k$ being duration of the $k_{th}$ period):

$$N_n(t) = N_1(T_1)e^{-\lambda(t-T_1)} + N_2(T_2)e^{-\lambda(t-T_1-T_2)} + \ldots + N_n(T_n) \qquad (4)$$

The latter equation allows us to determine the isotope production resulting from arbitrarily complex irradiation scenarios. Of particular interest is the build-up resulting from cyclic irradiation with the possibility of some leakage loss during each cycle. If $N_c$ represents the isotopes produced in a single cycle of length $\tau$, and (1-f) is the fraction of the total air volume that leaks out each cycle then the result after n cycles is described by a geometric series:

$$N(n\tau) = N_c f^{n-1} e^{-(n-1)\lambda\tau} + \ldots + N_c f e^{-\lambda\tau} + N_c = N_c \frac{\left(1 - f^n e^{-\lambda n \tau}\right)}{1 - f e^{-\lambda\tau}} \qquad (5)$$

In our case the air was exposed to two consecutive periods of steady but different beam power during the time prior to the measurements described. Moreover, the air circulation in the NuMI target chase results in a periodic exposure to beam which consists of an irradiation period while it passes through the target and horn regions, followed by a cool-down period while it is passing through the remainder of the volume. The production during the irradiation period is also cyclic due to the periodic structure of the beam.

To construct the corresponding build up equation we start with isotope production while the air is in contact with the beam. If $\tau_p$ is the length of a beam pulse, and $\tau_b$ is the length of beam cycle then the isotope production from a beam pulse is obtained using equation (2):

$$N_b = P\frac{\tau_b}{\tau_p}\lambda^{-1}\left(1-e^{-\lambda\tau_p}\right) \quad (6)$$

During the time that the air is exposed to the beam it will be irradiated with a sequence of beam pulses. Assuming constant beam power, we can describe the production that occurs each time the air passes through the target and horn region by setting $N_c = N_b$ and $\tau = \tau_b$ in equation (5) and defining $\tau_a = n\,\tau$ as the time during which the air is exposed to beam in an air circulation cycle:

$$N_a = P\frac{\tau_b}{\tau_p}\lambda^{-1}\frac{\left(1-e^{-\lambda\tau_p}\right)\left(1-e^{-\lambda\tau_a}\right)}{\left(1-e^{-\lambda\tau_b}\right)} \quad (7)$$

If $\tau_c$ is the time it takes for the air to circulate once through the entire volume, then the build-up over a period of time T is obtained using equation (5) again but this time we set $N_c = N_a$ and $\tau = \tau_c$ and allow for some constant leak rate f, between exposures:

$$N(T) = P\frac{\tau_b}{\tau_p}\lambda^{-1}\left(1-f^{T/\tau_c}e^{-\lambda T}\right)\frac{\left(1-e^{-\lambda\tau_p}\right)\left(1-e^{-\lambda\tau_a}\right)}{\left(1-e^{-\lambda\tau_b}\right)\left(1-fe^{-\lambda\tau_c}\right)} \quad (8)$$

Finally, we use the latter equation to combine the two running periods. If $T_1$ is the time spent running at beam power $W_1$ and $T_2$ is the time spent running at beam power $W_2$ then the total number of isotopes produced is given by the following expression:

$$N = P\lambda^{-1}\frac{\tau_b}{\tau_p}\left[\frac{W_1}{W_2}\left(1-f^{T_1/\tau_c}e^{-\lambda T_1}\right)f^{T_2/\tau_c}e^{-\lambda T_2}+\left(1-f^{T_2/\tau_c}e^{-\lambda T_2}\right)\right]\frac{\left(1-e^{-\lambda\tau_p}\right)\left(1-e^{-\lambda\tau_a}\right)}{\left(1-e^{-\lambda\tau_b}\right)\left(1-fe^{-\lambda\tau_c}\right)} \quad (9)$$

Accuracy of the latter expression was verified by means of comparisons with results of an independent computer modeling of certain build-up and cool-down sequences. The agreement between the modeling results and Eq. (9) for the full exposure scenario, described in the next section, for the most important nuclides—$^{11}$C, $^{13}$N and $^{41}$Ar— is 0.5% and better.

*2.3. Data and analysis*

Two independent activated air measurements were made on the air taken from the NuMI target chase. The samples were taken after 14 hours of steady beam running at 308 kW followed by a 3.75-hour period in which the beam was steady at 243 kW. The total volume of the air in the chase is about $5.07 \times 10^8$ cm$^3$ and the circulation rate is $1.18 \times 10^7$ cm$^3$/s. The air volume in the target and horn regions, where most of the air activation takes place, is $8.23 \times 10^7$ cm$^3$. This leads to the air circulation period of 42.96 seconds of which 6.97 seconds are spent in the target and horn volumes and nearby where the air is exposed to the incident and produced secondary particles. The leak rate from the total volume was measured to be $1.46 \times 10^5$ cm$^3$/s, which corresponds to a 1.24% loss per circulation cycle. The beam pulse length and cycle time, $\tau_p$ and $\tau_b$, are equal to $8.6 \times 10^{-6}$ and 1.75 seconds, respectively.

### 2.3.1. High Purity Germanium (HPGe) detector data

For this measurement, a thin-walled sample container (see Fig. 2) the size of a paint can ($3.91 \times 10^3$ cm$^3$), was connected to the supply and return line. It was then positioned in front of the HPGe at a known distance. Both the detector and the thin walled container were shielded with lead bricks. First a background measurement with this configuration was taken. Then a sample of the air from the target chase was collected, isolated by valving off the outlet and the inlet of the thin-walled container simultaneously. The HPGe measures the γ-rays emitted from the decay of radionuclides. The γ-rays from $^{41}$Ar decays produce a peak in the spectrum at 1293.64 keV whereas the other isotopes commonly produced in the air, such as $^{11}$C, $^{13}$N and $^{15}$O, all produce a positron that annihilates to produce a γ-peak at 511 keV.

The HPGe detector also provided an absolute measurement of the $^{41}$Ar activity. Using various correction factors to take into account the measurement geometry, decay of radioisotopes and detector efficiency, one can get the measured normalized production rate shown in Section 4.

### 2.3.2. Decay rate data

The decay rate data was obtained by isolating a sample of activated air in a thick-walled container (see Fig. 2) and counting the number of radiative decays detected by a thin windowed Geiger-Mueller (GM) inside the container, in short intervals (two seconds each) over a period of approximately 140 minutes. The data was used to find a composition of radionuclides that provides the best fit to the compound decay curve measured.

## 3. MARS15 physics model updates and calculations

### 3.1. Predictive features of contemporary Monte Carlo codes

Modern modelling tools—FLUKA and MARS and, in some cases, GEANT4—predict most of HEP and radiation quantities quite accurately being in good agreement with experimental data. For example, at well-defined conditions, they predict particle production and energy deposition within 5-20% compared to measured data, and prompt dose and DPA within 20-30%. At the same time, predictions of production of nuclides, their decay and transmutation—in particular, responsible for air activation—are less certain due to:
- complexity of the phenomenon with energies ranging from 120 GeV to a fraction of an eV for LBNF,
- imperfection of the models and nuclear databases in some energy intervals with data for some elements contradicting each other,
- uncertainties in geometry, air flow, irradiation/cooling profiles and exact composition of all the materials comprising the setup that is crucial for low-energy neutrons with resonant structure of cross-sections in the sub-MeV region.

Therefore, an agreement with 30-50% for nuclide production is considered as very good.

*3.2. Radionuclide production on light target nuclei at high energies and thermal neutron energies*

Initial calculations performed with the MARS15 code revealed that the predicted ratio of production rates of $^{11}$C and $^{13}$N differs significantly from that measured at similar conditions at other accelerators [7, 8]. Therefore, an update to the corresponding physical model was performed using the approach developed at CERN specifically for the air activation modeling at high energy accelerators [9, 10]. In particular, the production cross sections on light target nuclei and $^{41}$Ar were obtained from available experimental data that were analyzed, corrected and extrapolated using available systematics. The update was performed for incident protons, neutrons, and charged pions at energies from 10 MeV up to 10 TeV, and for neutrons from 14 MeV down to $2\times10^{-4}$ eV. Calculations, performed after the update, revealed that the ratio of production rates of $^{11}$C and $^{13}$N is improved and qualitatively agrees with the ratios measured at other accelerators. As an example, the production cross section of $^{13}$N on oxygen nuclei is shown in Fig. 4.

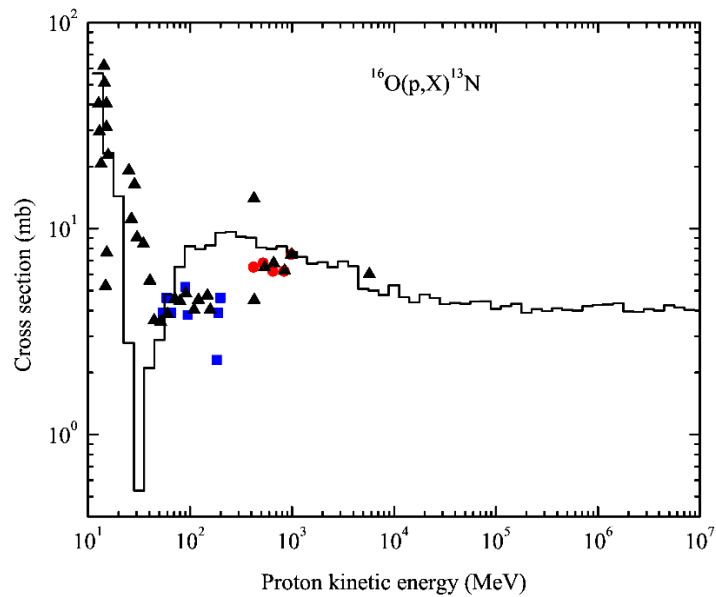

**Fig. 4.** Measured (symbols) and calculated (histogram) $^{13}$N production cross sections on oxygen nuclei for incident protons. The triangles, circles and squares correspond to data from Refs. [11], [12] and [13], respectively. The histogram is from Ref. [9].

*3.3. Event generator improvement for 1-200 MeV projectiles*

This development was motivated by the poor performance of intra-nuclear cascade and evaporation models (CEM and LAQGSM [14-15] in the MARS15 case) at projectile energies below a few tens of MeV (see Fig. 5). This typically resulted in an underestimation of low-energy neutron fluxes and, consequently, $^{41}$Ar production. A new inclusive/exclusive nuclear event generator based on the TENDL library [4] was created and implemented in the MARS15 base event generator at incident energies below 200 MeV for protons, neutrons, deuterons, tritons, $^{3}$He, $^{4}$He and gammas as projectiles. A mix-and-match procedure was applied to smooth the transition to the CEM and LAQGSM models used above 100 MeV. Various comparisons between the calculations and experimental data reveal a very good agreement (see Figs. 5-7).

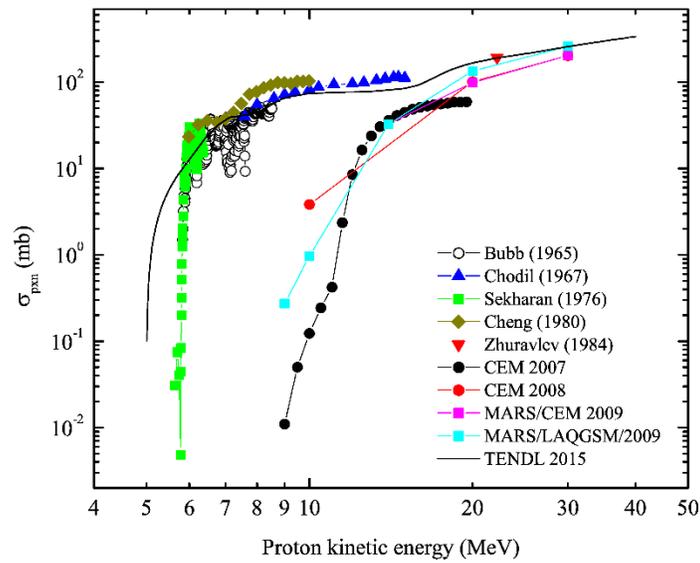

**Fig. 5.** Measured (the first five symbols) [16] and calculated (the last five symbols) [14-15] neutron production cross sections on aluminum nuclei.

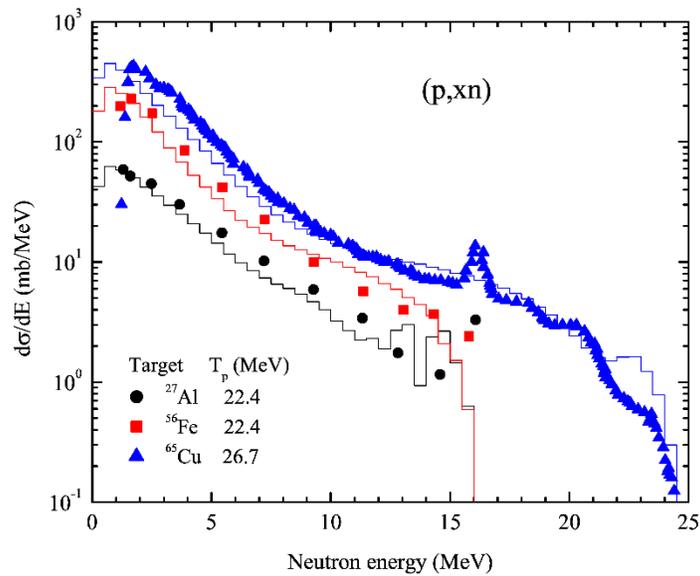

**Fig. 6.** Calculated with TENDL (lines) and measured (symbols) [16] energy distributions of secondary neutrons from different target nuclei due to incident protons of various kinetic energies ($T_p$).

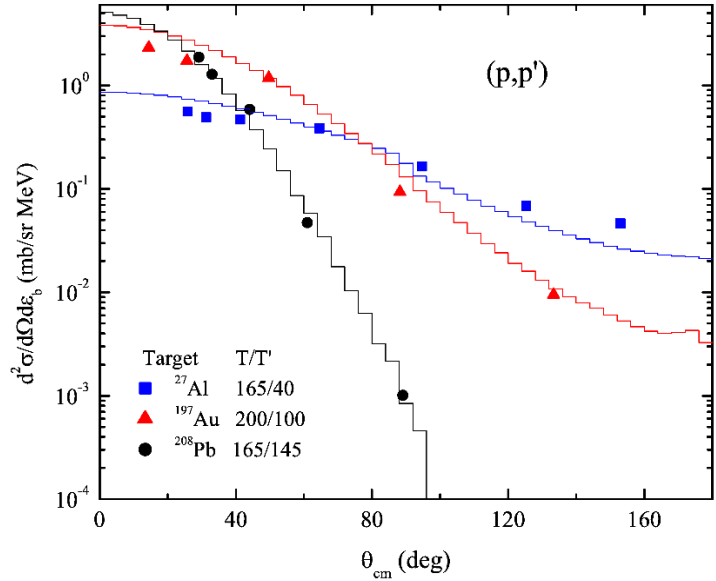

**Fig. 7.** Calculated with TENDL (lines) and measured (symbols) [17] angular distributions of secondary protons for light, medium and heavy target nuclei (projectile and ejectile energies, T and T´, respectively, are given in MeV).

Implementation of the TENDL data revealed that, all else being equal, the predicted production rate of $^{41}$Ar in the NuMI target chase improved by about 15%, compared to the previous prediction. The TENDL library also provides evaluated nuclear data separately for each stable isotope. This is an important feature which can serve as another reason for using the library at low energies, because reaction thresholds for different isotopes of the same element can differ significantly. Therefore, dealing with evaluated nuclear data files for each isotope instead of a single file for a corresponding natural mixture can provide more accurate results, for example, on nuclide production at low energies.

*3.4. Nuclide production*

Nuclide production in air regions is modelled in MARS15 in two complementary ways:
- Fully analog exclusive simulation of all nuclear interactions with scoring nuclides generated. Depending on the projectile energy E, the following event generators are used for that in the code: LAQGSM at E > 0.5 GeV; CEM at 0.1 < E < 0.5 GeV for all projectiles but neutrons and at 0.014 < E < 0.5 GeV for neutrons; at 0.001 < E < 0.1 GeV, TENDL for protons, d, t, $^3$He, $^4$He and photons, and CEM for all other projectiles; MCNP-like algorithms for neutrons at $10^{-12}$ GeV to 0.014 GeV.
- Energy-dependent hadron fluence folding with corresponding energy-dependent cross-sections in the course of Monte-Carlo modeling. The cross-sections are taken from Refs. [9] and [10].

Both methods used in the same run give similar results for the important nuclides, although not all the nuclides exist in the database of the second method. The first approach's outcome is noticeably more detailed, but requires substantially more CPU time to get statistical errors small enough for all the nuclides. Therefore, the second approach was used in these calculations for scoring nuclide production in the air.

*3.5. Simulation sequence*

A detailed description of the relevant geometry and material properties in the NuMI target chase have been implemented in the MARS simulation model. The air composition from Ref. [18] shown in Table 1 was used. The central segment of the MARS model is shown in Fig. 3. The entire model was built using ROOT-based geometry option recently implemented in MARS code [19]. At projectile energies above 8 GeV, projectile-target collisions are treated inclusively, while at lower energies the LAQGSM model is used. Energy cutoff applied to charged hadrons and neutrons was 100 keV and 0.001 eV, respectively. For precise simulation of low-energy neutron collisions, the calculations were performed in MCNP mode [3, 20].

**Table 1**

Atomic number densities of the most prominent elements and stable nuclides in the atmosphere at sea level at standard temperature and pressure [18].

| Element or isotope | % by volume | Isotopic abundance (%) | Atomic density (atom/cm$^3$) |
|---|---|---|---|
| Nitrogen (N$_2$) | 78.084 | | 4.1959×10$^{19}$ |
| $^{15}$N | | 0.364 | 1.5273×10$^{17}$ |
| $^{14}$N | | 99.636 | 4.1806×10$^{19}$ |
| Oxygen (O$_2$) | 20.9476 | | 1.1256×10$^{19}$ |
| $^{16}$O | | 99.757 | 1.1229×10$^{19}$ |
| $^{17}$O | | 0.038 | 4.2774×10$^{15}$ |
| $^{18}$O | | 0.205 | 2.3075×10$^{16}$ |
| Argon (Ar) | 0.934 | | 2.5094×10$^{17}$ |
| $^{36}$Ar | | 0.3336 | 8.3715×10$^{14}$ |
| $^{38}$Ar | | 0.0629 | 1.5784×10$^{14}$ |
| $^{40}$Ar | | 99.6035 | 2.4995×10$^{17}$ |

**4. Comparison between measurements and calculations**

One can see in Table 2 the NuMI data comparison with full MARS (DeTra) calculations for $^{41}$Ar, $^{11}$C, $^{13}$N and $^{15}$O. Nuclides $^{11}$C and $^{13}$N are produced mostly by high energy particles and they are easier to predict than $^{41}$Ar which is produced by low energy neutrons. Therefore, we were quite glad to obtain a 30% agreement on NuMI air activation between data and full Monte-Carlo calculations for $^{11}$C and $^{13}$N and better than a factor of two agreement for $^{41}$Ar. The uncertainties related to the measurements and MARS calculations (one standard deviation) are 12% and 15%, respectively.

**Table 2**

Measured and calculated production rate density (cm$^{-3}$ POT$^{-1}$ s$^{-1}$) for the most important radionuclides generated in the air in the beam enclosure of the NuMI target chase.

| Production rate | $^{41}$Ar | $^{11}$C | $^{13}$N | $^{15}$O |
|---|---|---|---|---|
| Measurement | 1.98×10$^{-12}$ | 6.38×10$^{-11}$ | 4.07×10$^{-11}$ | 3.50×10$^{-11}$ |
| MARS15 | 1.08×10$^{-12}$ | 4.44×10$^{-11}$ | 3.71×10$^{-11}$ | 4.16×10$^{-11}$ |
| MARS15/Measurement | 0.55 | 0.70 | 0.91 | 1.19 |

## 5. Conclusions

When comparing absolute production rates of $^{41}$Ar, the MARS prediction is 1.8 times lower than the measurements. Given a safety factor of 2 or 3 usually applied to code predictions, such an underestimate is quite acceptable. Taking into account the above-mentioned uncertainties associated with the measurements and MARS predictions, the agreement for the other significant radionuclides is good.

## Acknowledgements

This document was prepared using the resources of the Fermi National Accelerator Laboratory (Fermilab), a U.S. Department of Energy, Office of Science, HEP User Facility. Fermilab is managed by Fermi Research Alliance, LLC (FRA) acting under contract No. DE-AC02-07CH11359. Authors are grateful to B. Lundberg and S. D. Reitzner of Fermilab for providing an original computer model developed for the NuMI target chase.